\documentclass[twocolumn,showpacs,preprintnumbers,amsmath,amssymb]{revtex4}
\usepackage[dvips]{graphicx} % para incluir gráficos

\begin{document}
\title{Third quantization: modeling the universe as a 'particle' in a quantum field theory of the minisuperspace}

\author{S. J. Robles-P\'{e}rez}

\address{Instituto de F\'{\i}sica Fundamental, Consejo Superior de Investigaciones Cient\'{\i}ficas, Serrano 121, 28006 Madrid (SPAIN) and Estaci\'{o}n Ecol\'{o}gica de Biocosmolog\'{\i}a, Pedro de Alvarado 14, 06411 Medell\'{\i}n (SPAIN).}

\date{\today}

\begin{abstract}
The third quantization formalism of quantum cosmology adds simplicity and conceptual insight into the quantum description of the multiverse. Within such a formalism, the existence of squeezed and entangled states raises the question of whether the complementary principle of quantum mechanics has to be extended to the quantum description of the whole space-time manifold. If so, the \emph{particle} description entails the consideration of a multiverse scenario and the \emph{wave} description induces us to consider as well correlations and interactions among the universes of the multiverse.
\end{abstract}
\maketitle

\section{Introduction}

The so-called third quantization formalism was initially developed \cite{McGuigan1988, Rubakov1988, Strominger1990} to quantum mechanically represent the fluctuations of the space-time which, at the Planck length, provide it with a \emph{foam} structure \cite{Wheeler1957, Hawking1978}, in which virtual black holes, wormholes and baby universes cohabit. The main aim was to find an explanation for the vanishing value of the cosmological constant \cite{Rubakov1988, Coleman1988b}, which constituted the standard cosmological model at that time, and to study the effects that the space-time foam would produce in the coherence properties of matter fields \cite{Coleman1988, PFGD1990, PFGD1992b} as well as in the effective value of fundamental constants \cite{Giddings1988}.

On the other hand, a quantum description of the multiverse within the third quantization formalism has received some criticisms \cite{Vilenkin1994}. The most remarkable of them are that the scale factor cannot generally be taken as a time variable provided that it is not a monotonic function of time, and that the third quantization does not add anything to what it is already described within the usual quantum cosmological approach, whether this is given by the Wheeler-DeWitt equation or by the path integral formalism. However, as we shall try to briefly expose in this letter, it does add.

The scale factor cannot generally be taken as a time variable of the minisuperspace in a similar way as a well-defined time variable cannot be given in a general curved space-time. However, for universes with high degree of symmetry, the scale factor can be seen as a formal time-like variable of the minisupermetric that defines the geometrical structure of the corresponding minisuperspace. Furthermore, for large parent universes, the third quantization adds: i) simplicity of the model with respect to other approaches; ii) well-known standard procedures that can be applied to the quantum description of the universe; and, iii) conceptual insight of the quantum description of both  the universe and the multiverse. For instance, we shall see that the existence in the multiverse of quantum states having no classical analogue, like entangled and squeezed states, would no longer be associated to non-locality features because the concepts of locality or non-locality are meaningless in the quantum multiverse.

Nowadays, the multiverse has reached a wider acceptance \cite{Carr2007}, and the third quantization formalism supplies us with a quantum description of the multiverse that parallels that of a quantum field theory in a curved space-time. It also provides us with novel approaches for customary problems of quantum cosmology, like the boundary conditions, the problem of the cosmological constant, and the arrow of time, among others, and a framework in which new phenomena that were not contemplated so far may come out.

\section{Third quantization formalism}

The third quantization formalism consists of considering the wave function of the universe as a field to the quantized that propagates upon the minisuperspace. For instance, for a closed homogeneous and isotropic space-time minimally coupled to $n$ scalar fields, $\vec{\varphi}\equiv (\varphi_1, \ldots, \varphi_n)$, which represent the matter content of the universe, the Wheeler-DeWitt equation can be written as \cite{Kiefer2007}
\begin{equation}\label{WDWmini}
\left( -\frac{\hbar^2}{\sqrt{-G}} \partial_A (\sqrt{-G} G^{A B} \partial_B ) + \mathcal{V}(q^A) \right) \phi(q^A) = 0 ,
\end{equation}
where, $\{q^A\} \equiv \{a, \vec{\varphi}\}$, are the configuration variables of the minisuperspace, the potential $\mathcal{V}(q^A)$ is given by
$$
\mathcal{V}(q^A) = a^3 \Lambda - a + a^3 (V(\varphi_1) + \ldots + V(\varphi_n)) ,
$$
being $V(\varphi_i)$ the potential of the scalar field $\varphi_i$, and $G^{AB}$ is the inverse of the minisupermetric
\begin{equation}\label{minisuperM}
G_{AB} = {\rm diag}(-a, a^3, \ldots, a^3) ,
\end{equation}
with determinant, $G=-a^{3n+1}$. In Eq. (\ref{WDWmini}), $\phi(q^A)\equiv \phi(a, \vec{\varphi})$ is  the wave function of the universe being considered. The Lorentzian signature of the minisupermetric (\ref{minisuperM}) allows us to consider the scale factor as a formal time-like variable of the minisuperspace. This has not to be confused with a physical time variable in terms of 'clocks and rods'  measured by any observer. The relation between the scale factor and the Friedmann time, $t$, has to be found \emph{a-posteriori}.

We shall consider two particular cases: a massless scalar field in a de-Sitter space-time and a slow-varying field in a closed FRW space-time. For the massless scalar field, $\varphi$, the Wheeler-DeWitt equation (\ref{WDWmini}) explicitly reads
\begin{equation}
\ddot{\phi} + \frac{\dot{\mathcal{M}}}{\mathcal{M}} \dot{\phi} - \frac{1}{a^2} \phi'' + \omega^2(a) \phi = 0 ,
\end{equation}
where, $\phi\equiv \phi(a, \varphi)$ is the wave function of the universe, with $\dot{\phi}\equiv \frac{\partial \phi}{\partial a}$ and $\phi'\equiv \frac{\partial \phi}{\partial \varphi}$, $\dot{\mathcal{M}}\equiv \frac{\partial \mathcal{M}}{\partial a}$ with $\mathcal{M}=a$, and $\omega(a) = \frac{a}{\hbar}\sqrt{a^2 \Lambda - 1}$. Following the analogy between the third quantization formalism and a quantum field theory in a curved space-time, the wave function $\phi$ can be promoted to an operator that can be decomposed in normal modes as
\begin{equation}\label{decomposition}
\hat{\phi} = \frac{1}{\sqrt{2 \pi}} \int dk \left( e^{i k \varphi} A_{k}(a) \hat{c}_{k}  + e^{-i k \varphi} A^*_{k}(a) \hat{c}^\dag_{k} \right) ,
\end{equation}
where $\hat{c}_k$ and $\hat{c}^\dag_k$ are the usual annihilation and creation operators of the harmonic oscillator with mass and frequency $\mathcal{M}(a)$ and $\omega(a)$, respectively, evaluated on the boundary hypersurface $\Sigma_0$, for which $a=a_0$ and $\varphi=\varphi_0$. They represent annihilation and creation of universes, respectively, within the framework of the third quantization formalism. The probability amplitudes, $A_k(a)$ and $A^*_{k}(a)$, satisfy the equation of a damped harmonic oscillator with a mode-dependent frequency given by
\begin{equation}\label{frequency}
\omega_k = \frac{1}{\hbar} \sqrt{a^4 \Lambda - a^2 + \frac{\hbar^2 k^2}{a^2}} ,
\end{equation}
where the last term in the radicand of Eq. (\ref{frequency}) is a quantum correction that does not appear in the classical picture. For $k=0$, there is a Lorentzian region for values, $a > a_t \equiv \frac{1}{ \sqrt{\Lambda}}$, and an Euclidean region for values, $a_t >a>0$. The transition hypersurface $\Sigma_t\equiv \Sigma(a_t)$ corresponds then to the appearance of time \cite{Kiefer2007}. For values $k_m > k > 0$, where \cite{RP2011b} $k_m =\frac{4}{27 \hbar^2 \Lambda^2}$, before reaching the collapse the Euclidean instanton finds a new Lorentzian region (see Ref. \cite{RP2011b}). Then, following a reasoning that parallels that proposed in Refs. \cite{Barvinsky2006, Barvinsky2007a, Barvinsky2007b}, two instantons can then be joined forming a double instanton that would give rise to an entangled pair of universes with a composite quantum state given by \cite{RP2011b}
\begin{widetext}
\begin{equation}\label{EntInstanton}
\phi_{I,II} = \int dk \; \left( e^{ i k (\varphi_I +\varphi_{II})} A_{I,k} (a) A_{II,k}(a) \, \hat{b}_{I,k}^\dag \hat{b}_{II,k}^\dag + e^{ - i k(\varphi_I +\varphi_{II})} A_{I,k}^*(a) A_{II,k}^*(a) \, \hat{b}_{I,k} \; \hat{b}_{II,k} \right)  ,
\end{equation}
\end{widetext}
where $\varphi_{I,II}$ are  the scalar fields of each single universe, labelled by $I$ and $II$, respectively.

\section{The 'particle' definition of the universe}

Let us now consider the case of a slow-varying field in a closed FRW space-time, for which the Wheeler-DeWitt equation (\ref{WDWmini}) reads
\begin{equation}\label{SVfield}
\ddot{\phi} + \frac{\dot{\mathcal{M}}}{\mathcal{M}} \dot{\phi} + \omega_{sv}^2(a, \varphi) \phi = 0 ,
\end{equation}
where, $\omega_{sv}(a,\varphi) = \frac{a}{\hbar}\sqrt{a^2 V(\varphi) - 1}$. Eq. (\ref{SVfield}) is formally the equation of a harmonic oscillator defined on minisuperspace. Different representations can be chosen to quantum mechanically describe the universes. For instance, we could use the constant operators given in Eq. (\ref{decomposition}). However, in terms of such representation, the number operator, $\hat{N}_0 \equiv \hat{c}^\dag_{k} \hat{c}_{k}$, is not an invariant operator and its eigenvalues depend on the value of the scale factor. 

We would expect that, for a given model of the multiverse, the number of universes would not depend on the value of the scale factor of a particular single universe provided that we are not considering interaction terms in the Hamiltonian. Thus, it seems appropriate imposing the boundary condition that the number of universes of the multiverse is represented by an invariant operator. In the case of a slow-varying field, the Hamiltonian for which the Heisenberg equations of motion give rise to Eq. (\ref{SVfield}), is given by
\begin{equation}\label{Hamiltonian}
\hat{H} = \frac{1}{2 \mathcal{M}} \hat{P}_\phi^2 + \frac{\mathcal{M} \omega_{sv}^2}{2} \hat{\phi}^2 ,
\end{equation}
where $\hat{P}_\phi$ is the third quantized momentum conjugated to the wave function operator $\hat{\phi}$. Then, an invariant representation can be found by following the method developed by Lewis \cite{Lewis1969} and others \cite{Pedrosa1987, Dantas1992, Vergel2009}. It is given by the operators,
\begin{eqnarray}\label{b1}
\hat{b}(a) &\equiv& \sqrt{\frac{1}{2 \hbar}} \left( \frac{1}{R} \hat{\phi} + i (R \hat{P}_{\phi} - \dot{R} \hat{\phi}) \right) , \\ \label{b2}
\hat{b}^\dag(a) &\equiv& \sqrt{\frac{1}{2 \hbar}} \left( \frac{1}{R} \hat{\phi}  - i (R \hat{P}_{\phi} - \dot{R} \hat{\phi}) \right) , 
\end{eqnarray}
where,  $R=\sqrt{\phi_1^2 + \phi_2^2}$, being $\phi_1$ and $\phi_2$ two independent solutions of the Wheeler-de Witt equation (\ref{SVfield}). Then, $\hat{N} \equiv \hat{b}^\dag \hat{b}$, is an invariant operator fulfilling the boundary condition of the multiverse. However, in terms of the invariant representation, the Hamiltonian (\ref{Hamiltonian}) turns out to formally be the Hamiltonian of a parametric amplifier that, in quantum optics, is associated with the generation of entangled pairs of photons \cite{Scully1997}. Similarly, we can interpret that the Hamiltonian (\ref{Hamiltonian}) is associated, in the representation given by Eqs. (\ref{b1}-\ref{b2}), with the creation and annihilation of entangled pairs of universes in a multiverse scenario \cite{RP2011b}. 

However, for an observer inhabiting a large parent universe, for which $a\gg 1$, the appropriate representation of the universe is given by the asymptotic representation which is described in terms of the usual creation and annihilation operators of the harmonic oscillator with mass $\mathcal{M}(a)$ and frequency $\omega(a,\varphi)$. It turns out to be that \cite{RP2011b} the vacuum state of the multiverse becomes, in the asymptotic representation, a two-mode squeezed state that represents an entangled pair of universes. The quantum state of each single universe of the entangled pair turns out to be then given by a thermal state, which is indistinguishable from a classical mixture and whose thermodynamical properties can be computed. In particular, the entropy and energy of entanglement may provide us with a time variable and with a vacuum energy, respectively, for each single universe \cite{RP2011b, RP2012}.

\section{Wave-particle duality in the multiverse}

The existence of squeezed and entangled states in the multiverse raises the question of whether Bell's inequalities can be violated by such states in the multiverse, too. If so, however, such violation could not be related to non-locality features because there is no space-time among the universes of the quantum multiverse. It would be rather related to the interdependence of the quantum states that represent different universes.

The violation of classical inequalities by entangled and squeezed states in quantum optics is fundamentally related to the concept of complementary of the quantum theory \cite{Reid1986}. The extension of the complementary principle to the quantum multiverse would imply that the quantum state of the whole space-time manifold has to be complementary described in terms of \emph{particles and waves}. In terms of \emph{particles} means that, in an appropriate representation, the universes have to be considered as individual entities, giving rise to a multiverse scenario. In terms of \emph{waves} implies that interference effects among universes should be contemplated as well. Such a vision of the multiverse opens the door for novel approaches to customary problems in quantum cosmology. For instance, the thermodynamics of entanglement provides us with a new tool for studying the thermodynamical properties of a single universe and the related features of the vacuum energy and the arrow of time.

Furthermore, the consideration of the multiverse as a collective system also supplies us with unexpected phenomena. For instance, in Ref. \cite{Alonso2012} it is studied the quantum state of a multiverse made up of de-Sitter universes all of them with the same value, $\Lambda$, of their cosmological constants. Even considering no interaction among the universes, the energy spectrum of the collective system splits into two energy levels, which correspond respectively to two normal modes of the given representation, one of which corresponds to a state of a very small value of the energy. Then, the universes could initially be created in the excited level, which would provide us with a large value of the vacuum energy in the initial state of the universe, and decay afterwards into the ground state that entails a very small value of their effective cosmological constants. The effect is even more evident when a 'nearest' interaction is considered \cite{Alonso2012}. The spectrum splits then into a large number of energy levels, being the ground state a state with a very small value of the vacuum energy. The reasoning made in Ref. \cite{Alonso2012} is not conclusive but it plainly shows that the logic of the multiverse is rather different than the logic of a set of universes, being these taken individually, showing that the multiverse is much more that the mere sum of its parts.

\section*{Acknowledgments}

To the memory of Prof. Pedro F. Gonz\'alez D\'{\i}az, who always encouraged us to adopt a fearless and an open-minded attitude as well in science as in everyday life.

\bibliographystyle{apsrev}
\bibliography{bibliography}

\begin{thebibliography}{25}
\expandafter\ifx\csname natexlab\endcsname\relax\def\natexlab#1{#1}\fi
\expandafter\ifx\csname bibnamefont\endcsname\relax
  \def\bibnamefont#1{#1}\fi
\expandafter\ifx\csname bibfnamefont\endcsname\relax
  \def\bibfnamefont#1{#1}\fi
\expandafter\ifx\csname citenamefont\endcsname\relax
  \def\citenamefont#1{#1}\fi
\expandafter\ifx\csname url\endcsname\relax
  \def\url#1{\texttt{#1}}\fi
\expandafter\ifx\csname urlprefix\endcsname\relax\def\urlprefix{URL }\fi
\providecommand{\bibinfo}[2]{#2}
\providecommand{\eprint}[2][]{\url{#2}}

\bibitem[{\citenamefont{McGuigan}(1988)}]{McGuigan1988}
\bibinfo{author}{\bibfnamefont{M.}~\bibnamefont{McGuigan}},
  \bibinfo{journal}{Phys. Rev. D} \textbf{\bibinfo{volume}{38}},
  \bibinfo{pages}{3031} (\bibinfo{year}{1988}).

\bibitem[{\citenamefont{Rubakov}(1988)}]{Rubakov1988}
\bibinfo{author}{\bibfnamefont{V.~A.} \bibnamefont{Rubakov}},
  \bibinfo{journal}{Phys. Lett. B} \textbf{\bibinfo{volume}{214}},
  \bibinfo{pages}{503} (\bibinfo{year}{1988}).

\bibitem[{\citenamefont{Strominger}(1990)}]{Strominger1990}
\bibinfo{author}{\bibfnamefont{A.}~\bibnamefont{Strominger}}, in
  \emph{\bibinfo{booktitle}{Quantum Cosmology and Baby Universes}}, edited by
  \bibinfo{editor}{\bibfnamefont{S.}~\bibnamefont{Coleman}},
  \bibinfo{editor}{\bibfnamefont{J.~B.} \bibnamefont{Hartle}},
  \bibinfo{editor}{\bibfnamefont{T.}~\bibnamefont{Piran}}, \bibnamefont{and}
  \bibinfo{editor}{\bibfnamefont{S.}~\bibnamefont{Weinberg}}
  (\bibinfo{publisher}{World Scientific, London, UK}, \bibinfo{year}{1990}),
  vol.~\bibinfo{volume}{7}.

\bibitem[{\citenamefont{Wheeler}(1957)}]{Wheeler1957}
\bibinfo{author}{\bibfnamefont{J.~A.} \bibnamefont{Wheeler}},
  \bibinfo{journal}{Ann. Phys.} \textbf{\bibinfo{volume}{2}},
  \bibinfo{pages}{604} (\bibinfo{year}{1957}).

\bibitem[{\citenamefont{Hawking}(1978)}]{Hawking1978}
\bibinfo{author}{\bibfnamefont{S.~W.} \bibnamefont{Hawking}},
  \bibinfo{journal}{Nucl. Phys. B} \textbf{\bibinfo{volume}{144}},
  \bibinfo{pages}{349} (\bibinfo{year}{1978}).

\bibitem[{\citenamefont{Coleman}(1988{\natexlab{a}})}]{Coleman1988b}
\bibinfo{author}{\bibfnamefont{S.}~\bibnamefont{Coleman}},
  \bibinfo{journal}{Nucl. Phys. B} \textbf{\bibinfo{volume}{310}},
  \bibinfo{pages}{643} (\bibinfo{year}{1988}{\natexlab{a}}).

\bibitem[{\citenamefont{Coleman}(1988{\natexlab{b}})}]{Coleman1988}
\bibinfo{author}{\bibfnamefont{S.}~\bibnamefont{Coleman}},
  \bibinfo{journal}{Nucl. Phys. B} \textbf{\bibinfo{volume}{307}},
  \bibinfo{pages}{867} (\bibinfo{year}{1988}{\natexlab{b}}).

\bibitem[{\citenamefont{Gonz{\'a}lez-D{\'\i}az}(1990)}]{PFGD1990}
\bibinfo{author}{\bibfnamefont{P.~F.} \bibnamefont{Gonz{\'a}lez-D{\'\i}az}},
  \bibinfo{journal}{Phys. Rev. D} \textbf{\bibinfo{volume}{42}},
  \bibinfo{pages}{3983} (\bibinfo{year}{1990}).

\bibitem[{\citenamefont{Gonz{\'a}lez-D{\'\i}az}(1992)}]{PFGD1992b}
\bibinfo{author}{\bibfnamefont{P.~F.} \bibnamefont{Gonz{\'a}lez-D{\'\i}az}},
  \bibinfo{journal}{Phys. Rev. D} \textbf{\bibinfo{volume}{45}},
  \bibinfo{pages}{499} (\bibinfo{year}{1992}).

\bibitem[{\citenamefont{Giddings and Strominger}(1988)}]{Giddings1988}
\bibinfo{author}{\bibfnamefont{S.~B.} \bibnamefont{Giddings}} \bibnamefont{and}
  \bibinfo{author}{\bibfnamefont{A.}~\bibnamefont{Strominger}},
  \bibinfo{journal}{Nucl. Phys. B} \textbf{\bibinfo{volume}{207}},
  \bibinfo{pages}{854} (\bibinfo{year}{1988}).

\bibitem[{\citenamefont{Vilenkin}(1994)}]{Vilenkin1994}
\bibinfo{author}{\bibfnamefont{A.}~\bibnamefont{Vilenkin}},
  \bibinfo{journal}{Phys. Rev. D} \textbf{\bibinfo{volume}{50}},
  \bibinfo{pages}{2581} (\bibinfo{year}{1994}).

\bibitem[{\citenamefont{Carr}(2007)}]{Carr2007}
\bibinfo{editor}{\bibfnamefont{B.}~\bibnamefont{Carr}}, ed.,
  \emph{\bibinfo{title}{Universe or Multiverse}} (\bibinfo{publisher}{Cambridge
  University Press, Cambridge, UK}, \bibinfo{year}{2007}).

\bibitem[{\citenamefont{Kiefer}(2007)}]{Kiefer2007}
\bibinfo{author}{\bibfnamefont{C.}~\bibnamefont{Kiefer}},
  \emph{\bibinfo{title}{Quantum gravity}} (\bibinfo{publisher}{Oxford
  University Press, Oxford, UK}, \bibinfo{year}{2007}).

\bibitem[{\citenamefont{Robles-P{\'e}rez and
  Gonz{\'a}lez-D{\'\i}az}(2011)}]{RP2011b}
\bibinfo{author}{\bibfnamefont{S.}~\bibnamefont{Robles-P{\'e}rez}}
  \bibnamefont{and} \bibinfo{author}{\bibfnamefont{P.~F.}
  \bibnamefont{Gonz{\'a}lez-D{\'\i}az}}, \bibinfo{journal}{[arXiv:1111.4128]}
  (\bibinfo{year}{2011}), \eprint{arXiv:1111.4128}.

\bibitem[{\citenamefont{Barvinsky and Kamenshchik}(2006)}]{Barvinsky2006}
\bibinfo{author}{\bibfnamefont{A.~O.} \bibnamefont{Barvinsky}}
  \bibnamefont{and} \bibinfo{author}{\bibfnamefont{A.~Y.}
  \bibnamefont{Kamenshchik}}, \bibinfo{journal}{JCAP}
  \textbf{\bibinfo{volume}{0609}}, \bibinfo{pages}{014} (\bibinfo{year}{2006}),
  \eprint{hep-th/0605132}.

\bibitem[{\citenamefont{Barvinsky and Kamenshchik}(2007)}]{Barvinsky2007a}
\bibinfo{author}{\bibfnamefont{A.~O.} \bibnamefont{Barvinsky}}
  \bibnamefont{and} \bibinfo{author}{\bibfnamefont{A.~Y.}
  \bibnamefont{Kamenshchik}}, \bibinfo{journal}{J. Phys. A}
  \textbf{\bibinfo{volume}{40}}, \bibinfo{pages}{7043} (\bibinfo{year}{2007}),
  \eprint{hep-th/0701201}.

\bibitem[{\citenamefont{Barvinsky}(2007)}]{Barvinsky2007b}
\bibinfo{author}{\bibfnamefont{A.~O.} \bibnamefont{Barvinsky}},
  \bibinfo{journal}{Phys. Rev. Lett.} \textbf{\bibinfo{volume}{99}},
  \bibinfo{pages}{071301} (\bibinfo{year}{2007}), \eprint{0704.0083}.

\bibitem[{\citenamefont{Lewis and Riesenfeld}(1969)}]{Lewis1969}
\bibinfo{author}{\bibfnamefont{H.~R.} \bibnamefont{Lewis}} \bibnamefont{and}
  \bibinfo{author}{\bibfnamefont{W.~B.} \bibnamefont{Riesenfeld}},
  \bibinfo{journal}{J. Math. Phys.} \textbf{\bibinfo{volume}{10}},
  \bibinfo{pages}{1458} (\bibinfo{year}{1969}).

\bibitem[{\citenamefont{Pedrosa}(1987)}]{Pedrosa1987}
\bibinfo{author}{\bibfnamefont{I.~A.} \bibnamefont{Pedrosa}},
  \bibinfo{journal}{Phys. Rev. D} \textbf{\bibinfo{volume}{36}},
  \bibinfo{pages}{1279} (\bibinfo{year}{1987}).

\bibitem[{\citenamefont{Dantas et~al.}(1992)\citenamefont{Dantas, Pedrosa, and
  Baseia}}]{Dantas1992}
\bibinfo{author}{\bibfnamefont{C.~M.~A.} \bibnamefont{Dantas}},
  \bibinfo{author}{\bibfnamefont{I.~A.} \bibnamefont{Pedrosa}},
  \bibnamefont{and} \bibinfo{author}{\bibfnamefont{B.}~\bibnamefont{Baseia}},
  \bibinfo{journal}{Phys. Rev. A} \textbf{\bibinfo{volume}{45}},
  \bibinfo{pages}{1320} (\bibinfo{year}{1992}).

\bibitem[{\citenamefont{Vergel and Villase{\~n}or}(2009)}]{Vergel2009}
\bibinfo{author}{\bibfnamefont{D.~G.} \bibnamefont{Vergel}} \bibnamefont{and}
  \bibinfo{author}{\bibfnamefont{J.~S.} \bibnamefont{Villase{\~n}or}},
  \bibinfo{journal}{Ann. Phys.} \textbf{\bibinfo{volume}{324}},
  \bibinfo{pages}{1360} (\bibinfo{year}{2009}).

\bibitem[{\citenamefont{Scully and Zubairy}(1997)}]{Scully1997}
\bibinfo{author}{\bibfnamefont{M.~O.} \bibnamefont{Scully}} \bibnamefont{and}
  \bibinfo{author}{\bibfnamefont{M.~S.} \bibnamefont{Zubairy}},
  \emph{\bibinfo{title}{Quantum optics}} (\bibinfo{publisher}{Cambridge
  University Press, Cambridge, UK}, \bibinfo{year}{1997}).

\bibitem[{\citenamefont{Robles-P{\'e}rez}(2012)}]{RP2012}
\bibinfo{author}{\bibfnamefont{S.}~\bibnamefont{Robles-P{\'e}rez}},
  \bibinfo{journal}{[arXiv:1203.5774]}  (\bibinfo{year}{2012}),
  \eprint{arXiv:1203.5774}.

\bibitem[{\citenamefont{Reid and Walls}(1986)}]{Reid1986}
\bibinfo{author}{\bibfnamefont{M.~D.} \bibnamefont{Reid}} \bibnamefont{and}
  \bibinfo{author}{\bibfnamefont{D.~F.} \bibnamefont{Walls}},
  \bibinfo{journal}{Phys. Rev. A} \textbf{\bibinfo{volume}{34}},
  \bibinfo{pages}{1260} (\bibinfo{year}{1986}).

\bibitem[{\citenamefont{Alonso-Serrano
  et~al.}(2012)\citenamefont{Alonso-Serrano, Bastos, Bertolami, and
  Robles-P{\'e}rez}}]{Alonso2012}
\bibinfo{author}{\bibfnamefont{A.}~\bibnamefont{Alonso-Serrano}},
  \bibinfo{author}{\bibfnamefont{C.}~\bibnamefont{Bastos}},
  \bibinfo{author}{\bibfnamefont{O.}~\bibnamefont{Bertolami}},
  \bibnamefont{and}
  \bibinfo{author}{\bibfnamefont{S.}~\bibnamefont{Robles-P{\'e}rez}},
  \bibinfo{journal}{(in preparation)}  (\bibinfo{year}{2012}).

\end{thebibliography}

\end{document}